\documentclass[traditabstract, bibyear]{aa} 
\usepackage{graphicx}
\usepackage{txfonts}
\usepackage{color}
\usepackage{amstext}
\usepackage{answers}
\usepackage[table,x11names]{xcolor}

\begin{document}

   \title{On the hierarchical triple nature of the former red nova precursor candidate KIC~9832227}

   \author{Geza Kovacs\inst{1}
           \and 
           Joel D. Hartman\inst{2}
	   \and 
	   G\'asp\'ar \'A. Bakos\inst{2,} \inst{3,} \inst{4}}
	  
   \institute{Konkoly Observatory of the Hungarian Academy of Sciences, 
              Budapest, 1121 Konkoly Thege ut. 15-17, Hungary \\
              \email{kovacs@konkoly.hu}  
	      \and
              Department of Astrophysical Sciences, Princeton University, NJ 08544, USA 
	      \and
	      Packard Fellow 
	      \and
              MTA Distinguished Guest Fellow, Konkoly Observatory, Hungary
             }

   \date{Received June 28, 2019 / Accepted ...}


 
  \abstract
{We revisit the issue of period variation of the recently claimed red nova 
precursor candidate KIC~9832227. By using the data gathered during the main 
mission of the Kepler satellite, those collected by ground-based wide-field 
surveys and other monitoring programs (such as ASAS-SN), we find that the 
currently available timing data strongly support a model consisting of the 
known W~UMa binary and a distant low-mass companion with an orbital period 
of $\sim 13.5$ years. The period of the W~UMa component exhibits a linear 
period decrease with a pace of $(1.10\pm 0.05)\times10^{-6}$~days per year, 
within the range of many other similar systems. This rate of decrease is 
several orders of magnitude lower than that of V1309~Sco, the first (and 
so far the only) well-established binary precursor of a nova observed a few 
years before the outburst. The high-fidelity fit of the timing data 
and the conformity of the derived minimum mass of $(0.38\pm 0.02)$~M$_\odot$  
of the outer companion from these data with the limit posed by the spectroscopic 
non-detection of this component, are in agreement with the suggested 
hierarchical nature of this system.}

   
\titlerunning{The hierarchical system KIC~9832227}
\authorrunning{Kovacs, Hartman \& Bakos}
   \maketitle
%

%
%
\section{Introduction}\label{introduction}
\label{sect1}
Merging stars, degenerate objects and black holes have been thought 
for a long time to be responsible for many energetic phenomena in 
the Universe (e.g., for a subclass of Type I supernovae: 
Kuncarayakti~\cite{kuncarayakti2016}, Podsiadlowski, 
Joss \& Hsu et al.~\cite{podsiadlowski1992}, 
Truran \& Cameron~\cite{truran1971}). The discovery of the first 
black hole merger by the LIGO collaboration in 2015 
(Abbott et al.~\cite{abbott2016}) highlighted further the significance 
of the merger events. At the less energetic side, there is the recently 
discovered group of optical transients in between supernovae and novae. 
The group, annotated as Luminous Red Novae, consists of only a handful 
of objects, and they are suspected to be the result of low-mass stellar 
merging after the common envelope phase 
(e.g., Pastorello et al.~\cite{pastorello2019}, 
Pejcha et al.~\cite{pejcha2017}). Unlike in the case of black holes, 
for stars, during the precursor period of the actual merging event, 
there is a chance to spot potential mergers on the basis of the size 
of the period decrease. Unfortunately, nova events are rare, and 
their pre-discovery requires intensive monitoring. So far, there is 
only one object that has been clearly identified with this method. 
Thanks to the serendipitous overlap of the early OGLE survey area 
toward the Galactic Center with the position of Nova Scorpii 2008 
(Nakano, Nishiyama \& Kabashima~\cite{nakano2008}, 
Mason et al.~\cite{mason2010}), Tylenda et al.~(\cite{tylenda2011}) 
was able to identify V1309~Sco as a possible progenitor of the nova 
and trace back both the light curve and the variation of the orbital 
period. This has resulted in a wonderful match between the overall brightening 
and the dramatic period decrease, and thereby providing direct evidence 
for the first time of the viability of the binary connection with 
the red nova events (see also the possible settling of the system 
in a blue straggler state some $10$ years after the outburst -- 
Ferreira et al.~\cite{ferreira2019}). 
 
Stimulated by the case of V1309~Sco, Molnar et al.~(\cite{molnar2017}) 
investigated the W~UMa system KIC~9832227 by using their own data 
combined by those of the Kepler satellite and wide-field surveys 
from earlier epochs. They found that the eclipse timing variation 
was similar in nature to that of V1309~Sco, and, from the exponential 
trend they predicted a likely nova explosion in 2022. The study heavily 
relied on a single point from an earlier epoch, that was shown to suffer 
from an unfortunate timing typo (Socia et al.~\cite{socia2018}). 
This has cast doubt on the merger hypothesis, and, because of the 
poor observational coverage, vaguely suggested a different origin 
of the timing variation, including the presence of a tertiary 
component. The purpose of the present work is to elucidate further 
the status of KIC~9832227 by using both so far unpublished data 
and publicly available archival observations. This work not only 
confirms the result of Socia et al.~(\cite{socia2018}) on the 
lack of strong overall period change, but yields also valuable 
support to the hierarchical model of the system.  

%
%
\section{Datasets, derived O-C values}
\label{sect2}
In revisiting the timing properties of KIC~9832227, we searched for 
additional sources of data to: 
i) cover the crucial period prior to $T_0=$JD~2455000.0 more densely; 
ii) extend the time span past 2017; 
iii) fill in as densely as possible also other parts of the time 
interval covered by any data available (we found this latter step 
also necessary, since not all the data employed in the former 
studies are readily available at this moment).   

\begin{figure}
 \vspace{0pt}
 \centering
 \includegraphics[angle=-90,width=80mm]{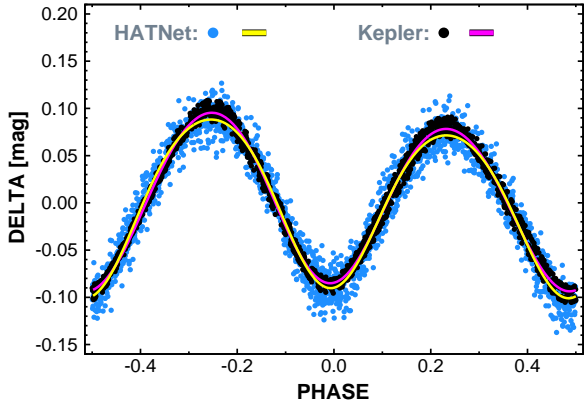}
 \caption{Comparison of the folded light curves and the corresponding 
          4-th order Fourier fits (continuous lines) for the HATnet and 
	  Kepler observations on KIC~9832227. We use the adjacent datasets 
          corresponding to $T_{\rm mid}=875$ and $878$, as given in 
          Table~\ref{o-c_data}. The ephemeris used is given in the same table. 
          To avoid crowdedness, only every 2-nd data points are plotted.}
\label{lc_comp}
\end{figure}

Importantly, the HATNet\footnote{\url{https://hatnet.org/}} survey 
(Bakos et. al.~\cite{bakos2004}) covered the Kepler field and we 
have a substantial amount of data available both in the pre- and 
post-$T_0$ periods. The earlier set from HATNet is between 
the observations made by the Vulcan telescope (Borucki et al.~\cite{borucki2001}, 
Socia et al.~\cite{socia2018}) and those gathered by Kepler in 
the beginning of the mission. We also have data obtained during 
the second part of the mission, establishing a good basis of 
comparison of the two datasets of highly different 
quality.\footnote{The photometric time series from 
the HatNet project is available at the CDS via anonymous 
ftp to cdsarc.u-strasbg.fr (130.79.128.5) or via
http://cdsarc.u-strasbg.fr/viz-bin/qcat?J/A+A/vol/page} Somewhat 
surprisingly, for the purpose of this study, the two datasets are 
quite compatible, as it is demonstrated in Fig.~\ref{lc_comp}. 
Interestingly, the depth of the primary minimum near phase zero 
is shallower than that of the secondary minimum. In other epochs 
this may change, likely because of stellar activity and binary 
interaction (Socia et al.~\cite{socia2018}). Because of these 
effects, we have a non-stationary light curve, yielding somewhat 
fuzzy segmented folded light curves also for the observations made 
by Kepler. For HATNet the higher observational noise further increases 
the scatter, but it is still tolerable for the goal of this work, 
i.e., deriving moments of the primary minima with an accuracy 
of $\sim 1$~min.    

%
%

Throughout this work we use light curves (LCs) obtained by simple 
ensemble photometry. For high-precision photometric time series 
analysis, application of filtering methods that clean up the data 
from colored noise play a crucial role in the full utilization of 
the signal content of the time series (i.e., see 
Kovacs et al.~\cite{kovacs2005}, 
Tamuz et al.~\cite{tamuz2005}, Bakos et al.~\cite{bakos2010}, 
Smith et al.~\cite{smith2012}). Our decision to stay with the 
more standard photometric data stems from the following reasons: 
i) Not all datasets contain these filtered LCs (e.g., ASAS-SN); 
ii) Since the amplitude of the light variation is large compared 
to that of the colored noise, the application of these filtering 
methods should consider this fact, otherwise the filtered signals 
will be distorted (see Kovacs~\cite{kovacs2018}). The systematics-filtered 
LCs, available at the download sites do not have a proper correction 
for this effect; 
iii) Tests made with systematics-filtered LCs as given at the download 
sites (i.e., not containing the signal reconstruction feature listed in 
ii) yield epochs within the error limits of the values derived from 
the ensemble photometric data. 

To further boost the coverage of the period variation, we made a 
thorough search among other survey projects, including space missions. 
This search ended up with data sources from the infrared survey satellite 
WISE (Wright et al.~\cite{wright2010}) through the ASAS-SN supernova search 
program (Shappee et al.~\cite{shappee2014} and Kochanek et al.~\cite{kochanek2017}) 
to the inventory of AAVSO. The detailed description of the datasets 
used is given in Appendix A. 

In dividing the data into segments of appropriate length to avoid 
smearing of the O-C values due to long individual time bases, we 
inspected the datasets separately and made a decision case-by-case. 
As a result, the time bases vary by several factors from one dataset 
to the other. Because of excessive errors, or improper phase coverage, 
certain observations -- although available -- were not used (e.g., 
data from the Palomar Transient Factory project -- see 
Rau et al.~\cite{rau2009} and Law et al.~\cite{law2009}). In the 
other extreme, for the Kepler data, we could go almost with a daily 
time base, but it would not serve any good for the purpose of the 
long-term study of our target. Nevertheless, it is worthwhile to 
note that such a study has certainly a relevance also from our point 
of view, since it is indicative of the inherent O-C jitter likely 
attributed to stellar activity. The analysis made by 
Socia et al.~(\cite{socia2018}) shows that the RMS of this physical 
scatter is about $1$--$2$~min on a time scale of several tens of 
days. As a result, for the Kepler data we use segments of $\sim 80$~d, 
and avoid mixing various quarters, if they show visible zero point 
differences. 

%
%
\begin{figure}
 \vspace{0pt}
 \centering
 \includegraphics[angle=-90,width=80mm]{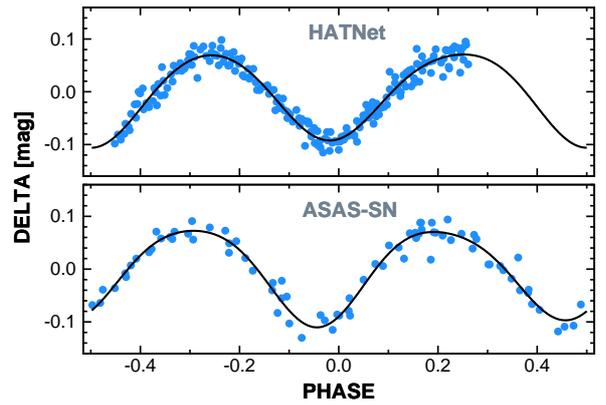}
 \caption{Examples of the performance of the fourth order Fourier fit 
          (continuous line) for partially and sparsely covered cases. 
	  The light curves shown correspond to the datasets of 
	  $T_{\rm mid}=998$ and $T_{\rm mid}=1840$ as given in 
	  Table~\ref{o-c_data}.}
\label{gapped_lc}
\end{figure}

It is important to choose a proper method to find the optimum way  
to estimate the epochs of the primary minima. Because of the varying 
light curve, this task is not entirely trivial, even though the basic 
light variation, originating from the eclipses of this W~UMa-type system, 
is decomposable into a Fourier sum of a few components only. 
We followed a simple empirical approach to select the best method. 
From the five proposed methods we selected the one that yielded the smallest 
residual scatter for the few-parameter O-C model to be described in detail 
in Sect.~\ref{sect3}. The following methods have been proposed  
(we also show the identifier of 
the O-C value corresponding to these methods as labelled in Table~\ref{o-c_data}). 
\begin{itemize}
\item
(a) {\em Fourier fit I -- {\rm OC1}:} Fourth-order Fourier fit to the 
entire phase-folded light curve (as given on a certain timebase) and 
search for the primary minimum by using the densely-sampled synthetic 
light curve. 
\item
(b) {\em Fourier fit II -- {\rm OC2}:} As in (a), but search for the 
moment of the secondary eclipse and shift the value by half of the 
period (we recall that the secondary eclipses are in many cases deeper 
than the primary eclipses, therefore, it is not entirely a bold idea 
to use these events to predict the moments of the primary eclipse 
-- with the assumption of circular binary orbit). 
\item
(c) {\em Fourier fit I+II -- {\rm OC3}:} This is just a simple average of the 
O-C values obtained by (a) and (b). 
\item
(d) {\em Template fit Ia -- {\rm OC4}:} Use some `representative' light curve 
as a template and perform a simple residual minimization on the 
full orbital phase by shifting this template relative to the target 
time series. If the template is properly chosen, the amount of 
shift will be characteristic of the overall shift of the light 
curve, perhaps depending in a less extent on the temporal change 
in its shape. The template is generated by the Fourier fits in 
running the minimum search by  method (a).  
\item
(e) {\em Template fit Ib -- {\rm OC5}:} As in (d), but the RMS of the residual 
is calculated only in a restricted range of the phase of the primary 
minimum (in our case this is $\pm 0.1$ in the units of the orbital 
period). 
\end{itemize}

In the O-C estimates based on Fourier fits and in the calculation of 
the templates for methods (d) and (e), the individual time series {X(i)} 
are fitted with a low order Fourier sum of 
%
%
\begin{eqnarray}
\label{sin4}
X(i) = A_0 + \sum_{j=1}^{m} A_{\rm j}\sin (2\pi j(t_{\rm i}-t_0)/P_0+\varphi_{\rm j}) \hskip 1mm ,
\end{eqnarray}  
where $t_0=2454953.949183$, $P_0=0.4579515$. This ephemeris, given 
by Socia et al.~(\cite{socia2018}) ensures to remain compatible with 
the earlier investigations and yield near zero O-C value at HJD$=2455000.0$. 
After experimenting with the Fourier order, we found that $m=4$ yields 
a good representation of all subsets, including even those with modest 
or gapped coverage of the full folded light curve (see Fig.~\ref{gapped_lc}). 

%
%
\begin{figure}
 \vspace{0pt}
 \centering
 \includegraphics[angle=0,width=75mm]{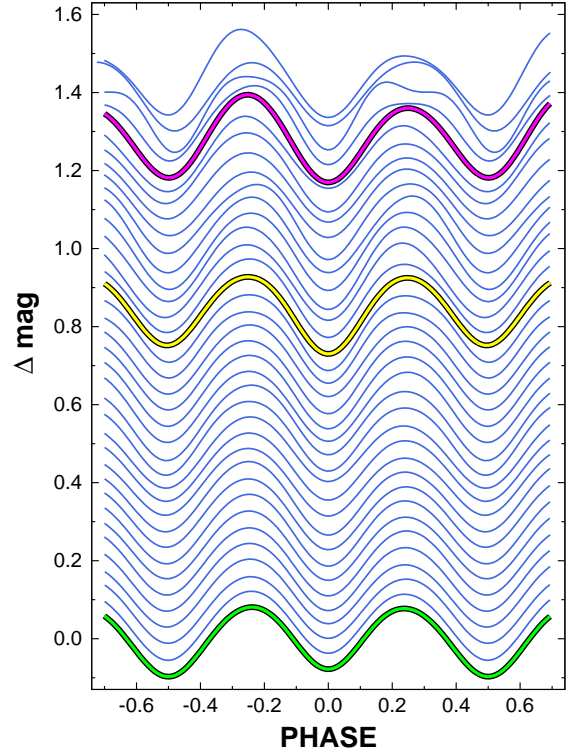}
 \caption{Fourier-fitted synthetic light curves of all the $37$ data 
          segments analyzed in this paper. The primary minima are 
	  shifted to zero phase. The time series are ordered on the 
	  basis of the closest neighbor as described in Sect.~\ref{sect2}. 
	  The pre-selected templates are shown by distinct colors. 
	  Following the notation of Table~\ref{o-c_data}, the green, 
	  yellow and magenta lines show the Fourier fits corresponding 
	  to segments {\em Kepler-4}, {\em AAVSO-V-1} and {\em AAVSO-V-4}, 
	  respectively.}
\label{template_lc}
\end{figure}

For O-C values of type (d) and (e), we need to select a `proper' 
template that yields the best fit for our low parameter model with 
the O-C values derived with this favorable template. We present 
the test leading to our final choice in Sect.~\ref{sect3}. Here 
we show only the Fourier fits to all the $37$ data segments analyzed 
in this paper and the three pre-selected candidates for master 
template (see Fig.~\ref{template_lc}). To aid the template testing 
-- i.e., to have an overall view on the light curve morphology when 
selecting some subset of the light curves -- the synthetic Fourier 
light curves have been ordered by the following algorithm. 

Starting with any of the light curves with some serial index 
$i\in [1,n]$, where $n$ is the number of light curves (in our case 
$37$), we find the `closest' neighbor to this light curve; i.e., 
we search for the light curve yielding the smallest RMS of the 
residuals obtained after subtracting the two time series from each 
other. We denote this RMS between the {\em i}-th test template and 
its nearest neighbor of serial index {\em j} by $\sigma (i,j)$. 
Then, this `closest neighbor' is searched for its closest neighbor 
(excluding of course all the previously found close neighbors). We 
continue the search until the last closest neighbor is found. The 
mutual RMS values \{$\sigma (i,k)$\} are examined and their variance 
is computed: 
$\sigma^2(i)={1 \over n}\sum_{k=1}^{n}(\sigma(i,k)-\langle\sigma(i,k)\rangle)^2$, 
where $\langle\sigma(i,k)\rangle={1 \over n}\sum_{k=1}^{n}\sigma(i,k)$. 
The degree of smoothness of the ordering derived from the above 
search for the {\em i}-th test template is then characterized by 
$\sigma(i)$. We repeat this whole process with the next starting 
template and do it for all possible starting templates. The smoothness 
parameters \{$\sigma(i)$\} for these orderings are examined, and the 
one with the smallest value is selected (with the associated starting 
template and ordering). 

By using this algorithm we obtained the light curves shown in 
Fig.~\ref{template_lc} with the set {\em Kepler-4} as the starting 
template. It is clear that most of the light curves are very similar 
to each other, and that there are only $\sim 4$ data segments that 
do not fit the overall smooth morphological sequence. These segments 
are {\em AAVSO-V-2}, {\em WISE-W12}, {\em HAT-5} and {\em WASP-1} 
(see the top four light curves in Fig.~\ref{template_lc}). 

We summarize the datasets used and the O-C values derived in 
Table~\ref{o-c_data}. The errors of the moments (eOC\#) of the 
mid-eclipses have been computed from a simple Monte Carlo test. 
In the case of the moments derived from Fourier fits for the 
given subset, the best fit Fourier signal is perturbed by a 
realization of a Gaussian noise with the standard deviation 
obtained by the fit to the real data. Then, this mock signal is 
fitted in the same way as the real data and the moment of the 
mid-eclipse is stored. The process is repeated thousand times 
and the standard deviation of the moments of these mid-eclipse 
values is considered as the error of the moment of the observed 
eclipse. For the eclipse times derived from template fitting 
the procedure is the same, except for the second step, when,  
instead of fitting a Fourier sum to the mock signal, we use a 
template. 

%
%
\begin{table*}[!t]
  \caption{Times of the primary minima for KIC~9832227.}
  \label{o-c_data}
  \setlength{\tabcolsep}{4pt}
  \scalebox{1.0}{
  \begin{tabular}{crrrcrcrcrcrcl}
  \hline\hline 
  T$_{\rm obs}(1)$ & T$_{\rm mid}$  &  $\Delta{\rm T}$  &  OC1  & eOC1 &  OC2  &  eOC2 &      OC3  &       eOC3 &      OC4  &      eOC4 &      OC5  &      eOC5 &   Source$^1$ \\
  (BJD$^1$)  &  (d)  &  (d) & (min)  & (min ) & (min)   & (min)   & (min)   & (min) & (min)  & (min) & (min)  & (min) & \\
  \hline
  1451.56499 & $ -3549 $ & $ 176 $ & $  41.5 $ & $ 2.7 $ & $  41.2 $ & $  3.1 $ & $  41.4 $ & $ 2.4 $ & $  38.2 $ & $ 1.3 $ & $  33.6 $ & $ 2.0 $ & {\em NSVS-12} \\
  2838.22900 & $ -2162 $ & $  45 $ & $  22.6 $ & $ 2.0 $ & $  22.6 $ & $  2.0 $ & $  22.6 $ & $ 2.0 $ & $  22.6 $ & $ 2.0 $ & $  22.6 $ & $ 2.0 $ & {\em Vulcan} \\
  3134.06377 & $ -1867 $ & $   5 $ & $  19.9 $ & $ 3.3 $ & $  19.4 $ & $  3.2 $ & $  19.7 $ & $ 2.6 $ & $  14.5 $ & $ 1.6 $ & $  13.2 $ & $ 3.1 $ & {\em WASP-1} \\
  3188.09813 & $ -1813 $ & $  35 $ & $  14.2 $ & $ 0.4 $ & $  12.9 $ & $  0.4 $ & $  13.6 $ & $ 0.3 $ & $  13.8 $ & $ 0.2 $ & $  12.5 $ & $ 0.4 $ & {\em HAT-1} \\
  3280.14596 & $ -1721 $ & $  65 $ & $  13.6 $ & $ 0.4 $ & $  13.4 $ & $  0.5 $ & $  13.5 $ & $ 0.4 $ & $  13.2 $ & $ 0.3 $ & $  11.9 $ & $ 0.4 $ & {\em HAT-2} \\
  3584.67982 & $ -1416 $ & $  49 $ & $   8.0 $ & $ 0.3 $ & $   7.4 $ & $  0.4 $ & $   7.7 $ & $ 0.3 $ & $   7.9 $ & $ 0.3 $ & $   7.3 $ & $ 0.4 $ & {\em HAT-3} \\
  4183.21603 & $  -817 $ & $ 299 $ & $  -1.1 $ & $ 3.2 $ & $   0.3 $ & $  4.4 $ & $  -0.4 $ & $ 3.0 $ & $  -0.0 $ & $ 2.0 $ & $   2.6 $ & $ 3.1 $ & {\em ASAS-IV} \\
  4263.81717 & $  -737 $ & $  33 $ & $   1.3 $ & $ 0.7 $ & $  -4.5 $ & $  0.7 $ & $  -1.6 $ & $ 0.6 $ & $  -2.6 $ & $ 0.4 $ & $   1.3 $ & $ 0.5 $ & {\em WASP-3} \\
  4647.12158 & $  -353 $ & $  41 $ & $  -0.1 $ & $ 0.4 $ & $  -3.4 $ & $  0.5 $ & $  -1.8 $ & $ 0.4 $ & $  -2.6 $ & $ 0.3 $ & $  -1.3 $ & $ 0.4 $ & {\em WASP-4} \\
  4994.24732 & $    -6 $ & $  40 $ & $  -2.3 $ & $ 0.5 $ & $  -5.9 $ & $  0.6 $ & $  -4.1 $ & $ 0.4 $ & $  -4.0 $ & $ 0.3 $ & $  -1.3 $ & $ 0.5 $ & {\em Kepler-1} \\
  5073.93207 & $    74 $ & $  38 $ & $  -0.6 $ & $ 0.5 $ & $  -6.6 $ & $  0.6 $ & $  -3.6 $ & $ 0.5 $ & $  -4.0 $ & $ 0.3 $ & $  -0.0 $ & $ 0.4 $ & {\em Kepler-2} \\
  5147.65928 & $   147 $ & $  34 $ & $  -4.9 $ & $ 0.5 $ & $  -2.0 $ & $  0.6 $ & $  -3.4 $ & $ 0.4 $ & $  -3.3 $ & $ 0.3 $ & $  -4.6 $ & $ 0.5 $ & {\em Kepler-3} \\
  5321.68220 & $   322 $ & $  37 $ & $  -2.9 $ & $ 0.5 $ & $  -3.1 $ & $  0.5 $ & $  -3.0 $ & $ 0.4 $ & $  -2.6 $ & $ 0.3 $ & $  -2.6 $ & $ 0.6 $ & {\em Kepler-4} \\
  5397.70233 & $   398 $ & $  38 $ & $  -2.7 $ & $ 0.2 $ & $  -5.6 $ & $  0.3 $ & $  -4.1 $ & $ 0.2 $ & $  -4.6 $ & $ 0.3 $ & $  -2.0 $ & $ 0.3 $ & {\em Kepler-5} \\
  5408.69213 & $   408 $ & $  91 $ & $  -4.1 $ & $ 2.2 $ & $   8.9 $ & $  3.2 $ & $   2.4 $ & $ 2.2 $ & $  -2.6 $ & $ 2.1 $ & $  -4.0 $ & $ 3.4 $ & {\em WISE-W12} \\
  5473.72165 & $   473 $ & $  37 $ & $  -3.6 $ & $ 0.4 $ & $  -4.4 $ & $  0.4 $ & $  -4.0 $ & $ 0.4 $ & $  -4.0 $ & $ 0.2 $ & $  -3.3 $ & $ 0.5 $ & {\em Kepler-6} \\
  5557.98516 & $   558 $ & $  46 $ & $  -2.9 $ & $ 0.5 $ & $  -7.8 $ & $  0.5 $ & $  -5.4 $ & $ 0.4 $ & $  -5.3 $ & $ 0.3 $ & $  -2.0 $ & $ 0.5 $ & {\em Kepler-7} \\
  5644.99354 & $   645 $ & $  40 $ & $  -6.4 $ & $ 0.3 $ & $  -6.9 $ & $  0.3 $ & $  -6.6 $ & $ 0.2 $ & $  -6.6 $ & $ 0.2 $ & $  -5.9 $ & $ 0.2 $ & {\em Kepler-8} \\
  5722.38858 & $   722 $ & $  37 $ & $  -4.6 $ & $ 0.3 $ & $ -11.0 $ & $  0.3 $ & $  -7.8 $ & $ 0.2 $ & $  -7.9 $ & $ 0.3 $ & $  -3.3 $ & $ 0.3 $ & {\em Kepler-9} \\
  5797.95109 & $   798 $ & $  38 $ & $  -3.9 $ & $ 0.2 $ & $  -9.0 $ & $  0.2 $ & $  -6.4 $ & $ 0.2 $ & $  -6.6 $ & $ 0.0 $ & $  -3.3 $ & $ 0.3 $ & {\em Kepler-10} \\
  5820.84780 & $   821 $ & $  27 $ & $  -5.1 $ & $ 0.3 $ & $  -8.9 $ & $  0.3 $ & $  -7.0 $ & $ 0.2 $ & $  -6.6 $ & $ 0.1 $ & $  -4.6 $ & $ 0.3 $ & {\em HAT-4a} \\
  5875.34399 & $   875 $ & $  39 $ & $  -5.2 $ & $ 0.2 $ & $  -9.0 $ & $  0.5 $ & $  -7.1 $ & $ 0.3 $ & $  -6.6 $ & $ 0.2 $ & $  -4.6 $ & $ 0.3 $ & {\em Kepler-11} \\
  5877.63486 & $   878 $ & $  28 $ & $  -3.6 $ & $ 0.3 $ & $  -8.7 $ & $  0.4 $ & $  -6.2 $ & $ 0.3 $ & $  -5.9 $ & $ 0.2 $ & $  -3.3 $ & $ 0.4 $ & {\em HAT-4b} \\
  5998.07105 & $   998 $ & $   1 $ & $ -10.8 $ & $ 1.9 $ & $   1.6 $ & $ 11.4 $ & $  -4.6 $ & $ 6.3 $ & $  -7.9 $ & $ 1.0 $ & $  -7.3 $ & $ 1.7 $ & {\em HAT-5} \\
  6140.03581 & $  1140 $ & $  32 $ & $ -11.1 $ & $ 0.3 $ & $ -14.6 $ & $  0.2 $ & $ -12.9 $ & $ 0.2 $ & $ -12.5 $ & $ 0.1 $ & $  -9.2 $ & $ 0.2 $ & {\em Kepler-12} \\
  6214.22294 & $  1214 $ & $  41 $ & $ -12.6 $ & $ 0.2 $ & $ -15.2 $ & $  0.2 $ & $ -13.9 $ & $ 0.2 $ & $ -13.8 $ & $ 0.1 $ & $ -10.6 $ & $ 0.3 $ & {\em Kepler-13} \\
  6298.48454 & $  1298 $ & $  43 $ & $ -14.7 $ & $ 0.2 $ & $ -17.8 $ & $  0.2 $ & $ -16.2 $ & $ 0.1 $ & $ -15.8 $ & $ 0.0 $ & $ -12.5 $ & $ 0.3 $ & {\em Kepler-14} \\
  6382.74892 & $  1383 $ & $  40 $ & $ -12.8 $ & $ 0.2 $ & $ -18.1 $ & $  0.2 $ & $ -15.5 $ & $ 0.1 $ & $ -15.2 $ & $ 0.2 $ & $  -9.9 $ & $ 0.1 $ & {\em Kepler-15} \\
  6840.23198 & $  1840 $ & $ 139 $ & $ -27.9 $ & $ 2.9 $ & $ -27.2 $ & $  3.4 $ & $ -27.6 $ & $ 2.6 $ & $ -28.4 $ & $ 1.7 $ & $ -25.1 $ & $ 3.3 $ & {\em ASAS-SN-1} \\
  7201.09114 & $  2201 $ & $ 132 $ & $ -37.4 $ & $ 5.3 $ & $ -39.4 $ & $  4.5 $ & $ -38.4 $ & $ 3.9 $ & $ -39.6 $ & $ 1.9 $ & $ -33.0 $ & $ 3.2 $ & {\em ASAS-SN-2} \\
  7336.64185 & $  2336 $ & $   7 $ & $ -41.7 $ & $ 0.3 $ & $ -44.7 $ & $  0.3 $ & $ -43.2 $ & $ 0.2 $ & $ -45.5 $ & $ 0.2 $ & $ -33.6 $ & $ 0.3 $ & {\em AAVSO-V-1} \\
  7569.73270 & $  2569 $ & $ 132 $ & $ -51.0 $ & $ 4.9 $ & $ -53.2 $ & $  3.7 $ & $ -52.1 $ & $ 3.2 $ & $ -52.1 $ & $ 1.7 $ & $ -39.6 $ & $ 4.2 $ & {\em ASAS-SN-3} \\
  7906.77350 & $  2907 $ & $  60 $ & $ -67.5 $ & $ 0.4 $ & $ -67.5 $ & $  0.4 $ & $ -67.5 $ & $ 0.4 $ & $ -67.5 $ & $ 0.4 $ & $ -67.5 $ & $ 0.4 $ & {\em MLO} \\
  7951.19236 & $  2951 $ & $  45 $ & $ -71.0 $ & $ 1.3 $ & $ -68.2 $ & $  1.4 $ & $ -69.6 $ & $ 1.1 $ & $ -70.6 $ & $ 0.8 $ & $ -60.7 $ & $ 1.9 $ & {\em AAVSO-V-2} \\
  7988.74625 & $  2989 $ & $ 187 $ & $ -68.3 $ & $ 4.7 $ & $ -75.8 $ & $  5.6 $ & $ -72.1 $ & $ 4.3 $ & $ -71.9 $ & $ 2.4 $ & $ -62.6 $ & $ 4.3 $ & {\em ASAS-SN-4} \\
  8076.66821 & $  3076 $ & $  71 $ & $ -75.1 $ & $ 2.9 $ & $ -77.2 $ & $  2.8 $ & $ -76.2 $ & $ 2.3 $ & $ -70.6 $ & $ 0.9 $ & $ -64.0 $ & $ 1.9 $ & {\em AAVSO-V-3} \\
  8225.95586 & $  3226 $ & $  47 $ & $ -81.7 $ & $ 1.1 $ & $ -82.0 $ & $  1.0 $ & $ -81.8 $ & $ 0.9 $ & $ -83.1 $ & $ 0.5 $ & $ -67.9 $ & $ 0.9 $ & {\em AAVSO-V-4} \\
  8338.60728 & $  3338 $ & $  36 $ & $ -88.4 $ & $ 0.6 $ & $ -86.3 $ & $  0.6 $ & $ -87.3 $ & $ 0.5 $ & $ -87.0 $ & $ 0.4 $ & $ -83.1 $ & $ 0.5 $ & {\em AAVSO-V-5} \\
  8354.63426 & $  3354 $ & $ 150 $ & $ -90.3 $ & $ 2.0 $ & $ -90.6 $ & $  2.0 $ & $ -90.4 $ & $ 1.7 $ & $ -85.7 $ & $ 0.9 $ & $ -83.1 $ & $ 2.0 $ & {\em AAVSO-I-1} \\
  \hline
\end{tabular}} 

$^1$All input time series are transformed to the BJD timebase. \\
$^2$See Appendix A for the details of the preparation of each data segment. 
\begin{flushleft}
\begin{minipage}{1.0\linewidth}
{\bf Notes:}{\small  
T$_{\rm obs}(1)$ : observed moment of the primary minimum minus $2450000.0$, 
derived from a $4$-th order Fourier fit; 
T$_{\rm mid}$ : center of the observational time span minus $2455000.0$; 
$\Delta{\rm T}$ : half range of the observational time span; 
OC\# : various observed minus calculated moments of minima, as described in Sect.~\ref{sect2};  
eOC\# : $1\sigma$ statistical error of OC\#. 
The calculated epochs of the primary minima are given by 
$C=2454953.949183+E\times0.4579515$, where $E$ is the closest integer cycle 
number for the given primary minimum. All O-C values resulted from the 
analysis presented in this paper, except for the values of Vulcan and MLO. 
For these we adopted the values by Socia et al.~(\cite{socia2018}). 
See Sect.~\ref{sect3} and Table~\ref{mod_per} for the template selection 
in computing OC4 and OC5.}
\end{minipage}
\end{flushleft}
\end{table*}
%

%
%
\begin{figure}
 \vspace{0pt}
 \centering
 \includegraphics[angle=-90,width=80mm]{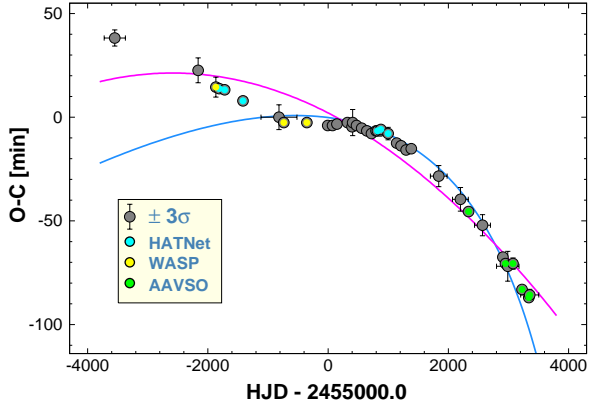}
 \caption{Observed minus calculated moments of the primary minima for 
          KIC~9832227. We used the template fit method yielding the 
	  OC4 values as described in Sect.~\ref{sect2}. The reference 
	  period and epoch of the primary eclipse are the same used 
	  by Socia et al.~(\cite{socia2018}), i.e., $P=0.4579515$~d, 
	  $EPOCH=2454953.949183$~BJD. Light blue line shows the exponential 
	  model of Molnar et al.~(\cite{molnar2017}) reaching singularity 
	  in $2022.2$ (i.e., at HJD$-2455000.0=4651$). Assuming only a 
	  linear period decrease, we get the best fitting parabola shown 
	  by the magenta line. Data associated with certain surveys are 
	  highlighted. See Table~\ref{o-c_data} for the full dataset.}
\label{inp_oc}
\end{figure}

The O-C data presented in Table~\ref{o-c_data} are plotted in 
Fig.~\ref{inp_oc}. It is worthwhile to emphasize that $3\sigma$ error 
bars are used, since standard $1\sigma$ bars would not be visible 
in most cases. It is also noted that former works by 
Molnar et al.~(\cite{molnar2017}) and Socia et al.~(\cite{socia2018}) 
did not use the WASP segment near ${\rm HJD}-2455000.0=-2000$, the HATNet, 
the ASAS-SN the AAVSO and the WISE data (but they used their own 
observations). The horizontal coordinate of each subset corresponds 
to the center of the time span of the given set. The horizontal bars 
are the half ranges of these time spans. 

We see that the Vulcan observations (second point from the left) and the 
HATNet and WASP data clearly breaks the exponential trend fitting based 
on data gathered after JD$=2455000.0$\footnote{In supporting this trend, 
there was also the point from the NSVS archive. However, as pointed out 
by Socia et al.~(\cite{socia2018}), this suffered from an MJD/HJD ambiguity 
in the work of Molnar et al.~(\cite{molnar2017}), that yielded the 
apparent confirmation of the exponential trend.}. Furthermore, assuming 
a simple linear period change for the approximation of the overall 
parabolic trend, leads also to an unsatisfactory fit (magenta line in 
Fig.~\ref{inp_oc}). The smooth, non-monotonic deviations from these 
simple representations indicate that the observed O-C values require a 
somewhat more involved modeling.

%
%
\section{Decomposition of the O-C variation}
\label{sect3}
Due to the limited number of data points, we aim for a modeling that 
is parsimonious with respect of the number of parameters fitted, and, at 
the same time, physically sound. This leads to the obvious choice of 
decomposing the observed O-C values into components of a linear period 
decrease and a periodic variation, modelled simply with a sine function 
of arbitrary amplitude and phase   
%
%
\begin{eqnarray}
\label{pol_sin_eq}
{\rm {O-C}} = c_1 + c_2t + c_3t^2 + c_4\sin\Big({2\pi t \over P_2}\Big) + c_5\cos\Big({2\pi t \over P_2}\Big) \hskip 1mm ,
\end{eqnarray}  
where O-C is the observed timing difference in [min], $t=HJD-2455000.0$ 
and $P_2$ is the period of the sinusoidal component. The other possibility 
would be to continue with higher order polynomial representation. To test 
briefly the goodness of fit for the pure polynomial approximation, we 
used the OC4 data (see Table~\ref{o-c_data} -- the observed moments of 
minima have been derived with the aid of template {\em Kepler-4}). 
Here, and throughout the paper, Eq.~\ref{pol_sin_eq} is fitted by 
simple Least Squares with equal weights. 
For the fourth order polynomial fit the unbiased RMS of the fit 
was $2.83$~min, exceeding significantly the fitting RMS of $1.38$~min 
of the pol$+$sin model. We may get similar RMS of $1.46$~min with the 
pure polynomial fit, 
if we increase the order 
by one, but then the fit suggests a steep downward trend for dates 
less than $-3600$. Further increase of the polynomial order leads to 
overshoot in the empty part of the dataset, between $-3600$ and 
$-2100$. We see that there is a preference toward the data representation 
indicated by Eq.~\ref{pol_sin_eq}. 

\begin{figure}
 \vspace{0pt}
 \centering
 \includegraphics[angle=-90,width=75mm]{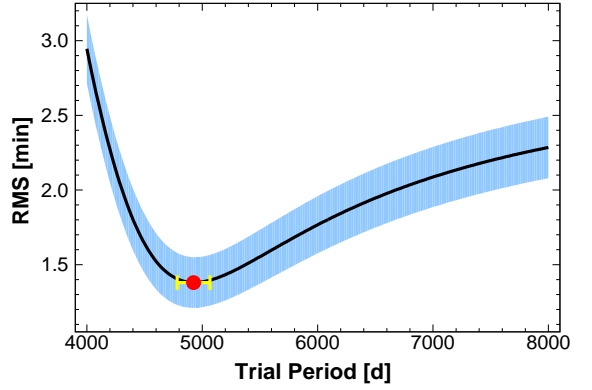}
 \caption{Period scan of the O-C data obtained by the template 
          fit method (see column OC4 of Table~\ref{o-c_data}). 
	  Equation~\ref{pol_sin_eq} is used to find the minimum 
	  RMS of the residuals (black line). The $1\sigma$ ranges 
	  of the residuals of the simple Monte Carlo simulations 
	  are shown by the shaded area. The optimum period and its 
	  $1\sigma$ error from the above simulations is shown by 
	  the red dot and its error bar.} 	  
\label{per_scan}
\end{figure}

To determine the best period for the sinusoidal component, we scanned 
the period in the $[4000,8000]$~d interval and searched for the lowest 
residual scatter given by the Least Squares fit of Eq.~\ref{pol_sin_eq}. 
With the best period found, we followed the same type of simple 
Monte Carlo method to estimate the error of the period, as described in 
Sect.~\ref{sect2} for the calculation of the errors of the O-C values. 
Figure~\ref{per_scan} shows the run of RMS for the finally accepted method, 
using full phase-folded light curve fit by employing the set {\em Kepler-4} 
as the template. Comparison with the other methods and templates is given 
in Table~\ref{mod_per}. 
We see that the full-phase template fit performs well, with some 
dependence on the template used. Based on these tests, 
we set the long-term modulation period $P_2$ equal to $(4925 \pm 142)$~d.    

%
%
%

%
%
\definecolor{magicmint}{rgb}{0.67, 0.94, 0.82}
\definecolor{almond}{rgb}{0.94, 0.87, 0.8}
\definecolor{blizzardblue}{rgb}{0.67, 0.9, 0.93}
\definecolor{paleaqua}{rgb}{0.74, 0.83, 0.9}
\begin{table}[!h]
  \caption{Modulation periods derived by various methods.}
  \label{mod_per}
  \scalebox{1.0}{
  \begin{tabular}{cccl}
  \hline\hline
   OC$\#$ & $P_2$ (d) & $\sigma_{\rm fit}$ (min) & Template \\
  \hline
  1  & $ 5427 \pm 266 $ & $ 1.684 $ & -- \\
  2  & $ 5045 \pm 492 $ & $ 3.678 $ & -- \\
  3  & $ 5206 \pm 243 $ & $ 1.852 $ & -- \\
  4  & $ 4965 \pm 185 $ & $ 1.633 $ & {\em AAVSO-V-4} \\
\rowcolor{paleaqua} & $ 4925 \pm 142 $ & $ 1.380 $ & {\em Kepler-4} \\
     & $ 4985 \pm 169 $ & $ 1.537 $ & {\em AAVSO-V-1} \\
  5  & $ 4683 \pm 276 $ & $ 3.261 $ & {\em AAVSO-V-4} \\
     & $ 5990 \pm 747 $ & $ 2.689 $ & {\em Kepler-4} \\
     & $ 4965 \pm 320 $ & $ 3.019 $ & {\em AAVSO-V-1} \\
\hline
\end{tabular}}
\begin{flushleft}
\begin{minipage}{1.0\linewidth}
{\bf Notes:}{\small  
The type of the fit employed on the light curves is given in the first 
column and defined in Sect.~\ref{sect2} (the first three refer to 
Fourier-based O-C values, the 4-th and 5-th to template-based values 
with the template indicated in the last column). Equation~\ref{pol_sin_eq} 
has been fitted by Least Squares to derive $\sigma_{\rm fit}$ for the 
best modulation period. The $1\sigma$ errors of the periods were calculated 
with the aid of a simple Monte Carlo method. The best solution is highlighted.}
\end{minipage}
\end{flushleft}
\end{table}

\begin{figure}
 \vspace{0pt}
 \centering
 \includegraphics[angle=0,width=75mm]{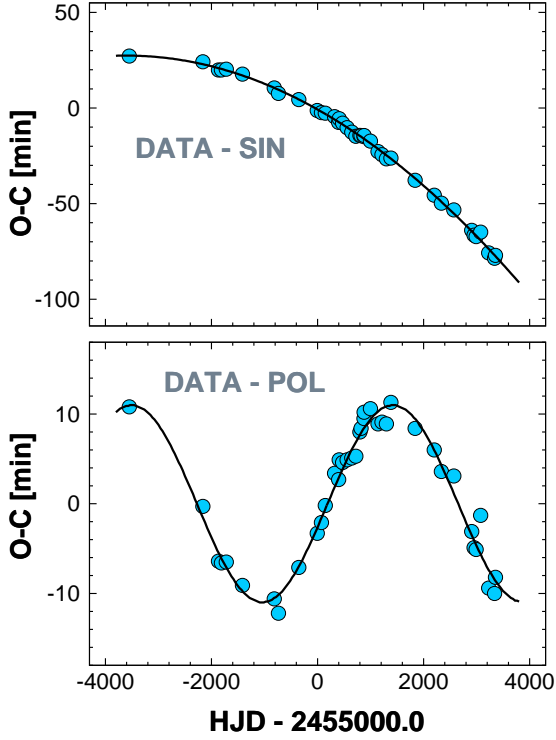}
 \caption{Result of the joint fit of a second order polynomial and a single 
          sinusoidal to the data shown Fig.~\ref{inp_oc}. The data plotted 
	  in the upper panel are free from the sinusoidal, whereas those 
	  shown in the lower panel from the polynomial component. See 
	  Table~\ref{fit_par} for the regression coefficients of these 
	  components.}
\label{pol_sin}
\end{figure}

With the period known, we can decompose the data into a quadratic and 
a periodic part. The result is shown in Fig.~\ref{pol_sin}. It seems 
that this simple model fits the data near the noise level expected 
from the physical jitter of the minima due to stellar activity 
(Socia et al.~\cite{socia2018}) and from the overall statistical error 
due to observational noise. 

%
%
%

%
%
\begin{table}[!h]
  \caption{Parameters of the polynomial$+$sinusoidal fit.}
  \label{fit_par}
  \scalebox{1.0}{
  \begin{tabular}{cll}
  \hline\hline
   Coeff & Value & Error  \\ 
 \hline
$c_1$  & $          -0.76257\times 10^{+0} $ & $ 0.38377\times 10^{+0} $ \\ 
$c_2$  & $          -0.15618\times 10^{-1} $ & $ 0.14628\times 10^{-3} $ \\ 
$c_3$  & $          -0.21632\times 10^{-5} $ & $ 0.09212\times 10^{-6} $ \\ 
$c_4$  & $\phantom{-}0.10659\times 10^{+2} $ & $ 0.35694\times 10^{+0} $ \\ 
$c_5$  & $          -0.26196\times 10^{+1} $ & $ 0.47482\times 10^{+0} $ \\ 
\hline
\end{tabular}}
\\
  \scalebox{1.0}{
  \begin{tabular}{llll}
  \multicolumn{4}{c}{Associated quantities} \\
  \hline
  Name         & Unit   &  Value               & Error \\ 
  \hline
  $P_2$        & [d]    &  $\phantom{-}4925  $ & $142$ \\
  RMS          & [min]  &  $\phantom{-}1.380 $ & $0.169$ \\
  {\em Pdot}   & [d/yr] &  $          -1.097\times 10^{-6}$ & $0.047\times 10^{-6}$ \\
  {\em Ampl}   & [min]  &  $ 10.976          $ & $0.365$ \\
  \hline
\end{tabular}}
\begin{flushleft}
\begin{minipage}{1.0\linewidth}
{\bf Notes:}{\small \  
For the meaning of the regression coefficients \{$c_i$\} see Eq.~\ref{pol_sin_eq}. 
Errors on \{$c_i$\} do not include the contribution from the period. 
}
\end{minipage}
\end{flushleft}
\end{table}

The regression parameters and their standard errors are listed in 
Table~\ref{fit_par}. The errors are $1\sigma$ standard deviations 
and derived directly from the inverse of the normal matrix. 
The error of the amplitude of the periodic component results 
from random simulations assuming independent Gaussian errors with 
standard deviations given on $c_4$ and $c_5$ from the normal matrix. 
Both Fig.~\ref{pol_sin} and the errors of the physically interesting 
quantities (period change {\em Pdot} and amplitude of the periodic 
component {\em Ampl}) together with the low residual scatter indicate 
that the chosen mathematical representation is quite suitable for  
the description of the available data. For a sanity check of 
the solution above, in Appendix B we present the solution obtained 
by the Markov Chain Monte Carlo (MCMC) method.  

%
%
%
\begin{figure}
 \vspace{0pt}
 \centering
 \includegraphics[angle=-90,width=80mm]{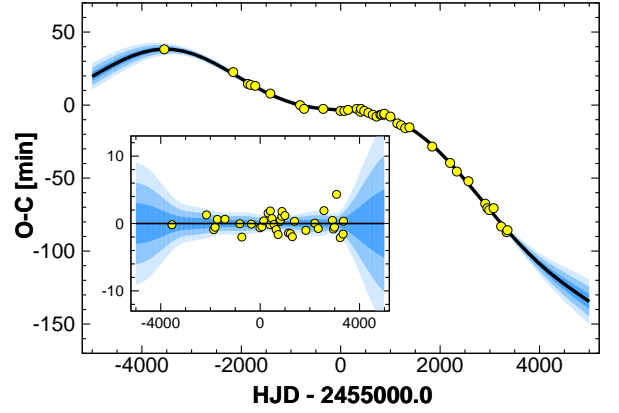}
 \caption{Polynomial$+$sinusoidal fit (thick black line) of the 
          observed O-C values (yellow dots, OC4 of Table~\ref{o-c_data}). 
	  In the order of lighter shading, the blue regions show 
	  the $1$, $2$ and $3$ sigma limits of the predicted O-C 
	  ranges, assuming errors in all estimated parameters, 
	  including $P_2$, the period of the long-term variation. 
	  The inset shows the deviation from the best fit to the 
	  currently available data.}
\label{oc_pred}
\end{figure}

In spite of the good fit to the observations, the predictive power 
of the derived polynomial$+$sinusoidal model is not very high. For 
a fixed period of the sinusoidal component one can give an estimate 
on the variance of the predicted O-C values for arbitrary moments 
by using the inverse of the normal matrix of the linear regression 
of Eq.~\ref{pol_sin_eq} 
%
%
\begin{eqnarray}
\label{pol_sin_pred}
\sigma_{\rm pred}^2(i) = \sigma_{\rm fit}^2 \sum_{j=1}^5 \sum_{k=1}^5 G(j,k)x(j,i)x(k,i) \hskip 1mm ,
\end{eqnarray}  
where $\sigma_{\rm fit}$ is the unbiased standard deviation of the 
residual of the regression, \{$G(j,k)$\} is the inverse of the normal 
matrix, and  \{$x(j,i)$\} is the set of vectors associated with the 
$j$-th regression coefficient $c_j$ (see Eq.~\ref{pol_sin_eq}) at 
moment $t_i$. Although the above formula is exact, it lacks accounting 
for the error introduced by the period $P_2$. To consider the contribution 
of the period estimation, we need to perform a Monte Carlo simulation. 
Here we use the synthetic data obtained by the fit to the original 
input time series with the best period. We add observational noise 
(Gaussian, with the standard deviation given by $\sigma_{\rm fit}$) 
and then fit the data by Eq.~\ref{pol_sin_eq}, where $P_2$ is also 
perturbed with a Gaussian error of $\sigma=142$ (see Table~\ref{fit_par}). 
We repeat this process $500$ times and arrive to the point-by-point 
standard deviations of the realization-dependent predicted O-C 
values. Figure~\ref{oc_pred} shows the resulting regions visited by 
the realizations, scaled by $\sigma_{\rm pred}$. We note that using 
Eq.~\ref{pol_sin_pred} (i.e., neglecting the effect of the $P_2$ 
ambiguity), we get a somewhat more restricted prediction space, with 
a $3\sigma$ limit corresponding to the $2\sigma$ limit of the 
result shown. In any case, if the model is correct, in the following 
$3$--$4$ years the periodic component will pass the minimum and 
starts to rise by a few minutes, that could be verified rather 
easily even by sporadic observations.  

%
%
\section{Physical interpretation}
\label{sect4}
First we discuss the possible physical causes giving rise to the periodic 
component of the O-C variation. There could be three possible 
causes of this type of variation: 
(i) change in the stellar structure due to stellar magnetic cycles 
(the so-called Applegate-effect -- see Applegate~\cite{applegate1992}); 
(ii) apsidal motion mainly due to the finite size and non-spherical shape 
of the components (e.g., Mazeh~\cite{mazeh2008}); 
(iii) light-time effect, due to an external perturber (e.g., 
Wolf~\cite{wolf2014}). 

\begin{figure}
 \vspace{0pt}
 \centering
 \includegraphics[angle=-90,width=80mm]{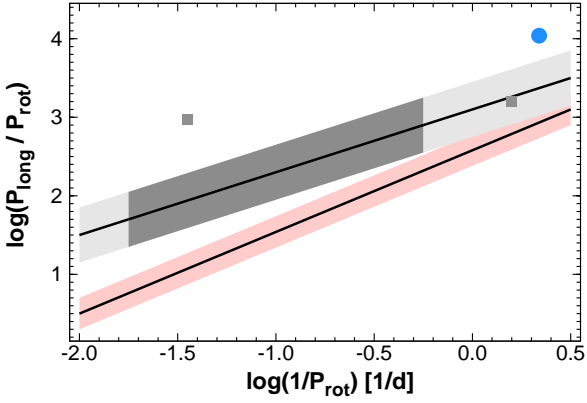}
 \caption{Position of KIC~9832227 (light blue dot) on the schematic  
          diagram relating the rotation frequency to the ratio of the 
	  period of the long-term cyclic luminosity variation due to 
	  stellar activity to the rotation period. The shaded area 
	  cover the data points shown by Almeida et al.~(\cite{almeida2019}), 
	  based on the studies of Vida et al.~(\cite{vida2013}, 
	  \cite{vida2014}), Ol\'ah et al.~(\cite{olah2009}), and 
	  Savanov~(\cite{savanov2012}). Pink shade is used for red dwarfs, 
          dark gray for other stars, with the extension to non-populated 
	  regions (light gray). Isolated small gray squares indicate 
	  objects distinct from the bulk of the sample.}
\label{applegate}
\end{figure}

For a brief check of possibility (i) we follow Almeida et al.~(\cite{almeida2019}) 
and use the empirical studies of Ol\'ah et al.~(\cite{olah2009}), 
Savanov~(\cite{savanov2012}) and Vida et al.~(\cite{vida2013}, \cite{vida2014}) 
to see if our target follows the relation between the rotation 
frequency and the magnetic cycle length. We make the obviously 
non-stringent assumption that the orbital period is a good proxy 
of the rotation period. The result is shown in Fig.~\ref{applegate}. 
Because our target has rather small error bar even on the longer 
period -- yielding $\sigma (log(P_{\rm long}/P_{\rm rot}))\sim 0.013$ -- 
the outlier status of this object is quite likely, assuming that 
the observed trend continues also in the currently poorly explored 
parameter regime. 

%
%
%

Possibility (ii) may occur due to the slow drift of the apsis line 
in systems with eccentric orbits. One of the characteristics of this 
phenomenon is the anticorrelation of the timings of the primary and 
secondary minima (e.g., Wolf, Zejda \& de Villiers~\cite{wolf2008}). 
For our target, this explanation is also unlikely. First, the eccentricity 
must be very small, as suggested by the good fit of the radial velocity 
data with the assumption of zero eccentricity (see Molnar et al.~\cite{molnar2017}). 
Second, the difference between the primary and secondary O-C values 
(see Table~\ref{o-c_data}) also supports the assumption of near zero eccentricity. 
Indeed, leaving out the secondary eclipse outliers {\em WISE-W12} and {\em HAT-5}, 
the average difference ($\Delta T = T_2-T_1-P_{\rm orb}/2$) is $-2.0$~min 
with a standard deviation of $2.7$~min for the remaining $37$ points. 
This implies -- see, e.g., Winn~\cite{winn2010} --  
$e\cos \omega ={\pi\Delta T \over 2P_{\rm orb}}=0.005\pm0.001$, where 
the error has been estimated from the error of the average $\Delta T$. 
Third, the period of the sinusoidal component of the O-C variation of 
our target is relatively short, and therefore, not too common among 
objects showing apsidal motion (see Hong et al.~\cite{hong2016} and 
Zasche et al.~\cite{zasche2018}). 

Possibility (iii) implies that the system is a hierarchical triplet, 
with an outer component on a $4925$~d orbit. The mass function can 
be estimated from the amplitude of the sinusoidal component of the 
total O-C variation. With the assumption of zero eccentricity, this 
yields: $M_3 \geq (0.38\pm 0.02)$~M$_\odot$, where the error is the 
standard deviation of the values obtained by including the Gaussian 
errors of all constituent parameters. We recall that this mass 
(although it is a lower limit) is still allowed by the test performed 
on the composite spectral broadening function by Molnar et al.~(\cite{molnar2017}). 
They suggest an upper limit of $0.5$~M$_\odot$ for any possible 
third component, assuming that it is a main sequence star. 

\begin{figure}
 \vspace{0pt}
 \centering
 \includegraphics[angle=-0,width=70mm]{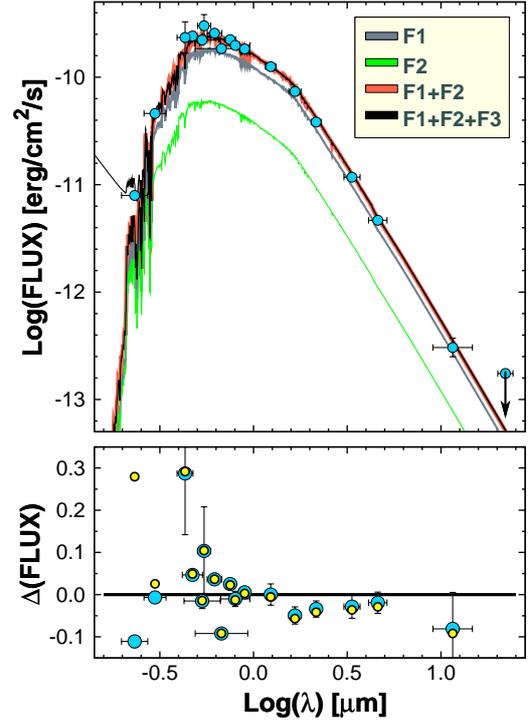}
 \caption{Comparison of the observed and theoretical SEDs for 
          KIC~9832227. Dots denote observed values, with $3\sigma$ 
	  vertical error bars, and equivalent waveband widths 
	  (horizontal bars). The inset shows the various theoretical 
	  SEDs: F1 - primary only, F2 - secondary only, F1$+$F2 -  
	  binary only, F1$+$F2$+$F3 - binary, with the hypothetical 
	  3rd component -- K/M dwarf or white dwarf (black lines, 
	  differ only in the UV). The lower panel shows the residual 
	  in respect of the three-component solution:  blue dots - 
	  white dwarf, yellow dots: K/M dwarf for the 3rd component. 
	  The stellar atmosphere models of 
	  Castelli, Gratton \& Kurucz~(\cite{castelli1997}) and 
	  Koester~(\cite{koester2010}) are used (see text and Appendix C   
	  for more details).}
\label{sed}
\end{figure}

Although it is not expected that the third star with its low luminosity 
(implied by its mass) will give any appreciable contribution to the total 
spectral energy distribution (SED) of the system, it is still useful 
to check the actual values and explore the effect of components different 
from a main sequence star. We gathered $18$ wide-band color values from 
the near ultraviolet to infrared regime. After converting all magnitudes 
to the Vega system, we compared these values by the theoretical SEDs 
obtained from various stellar atmosphere models (we refer to Appendix C   
for additional details on the data and the models). The final SED values 
are plotted in Fig.~\ref{sed}.   

%
%

Assuming that the third component is a main sequence star (a late 
K dwarf or an early M dwarf) we set its radius equal to $0.4$~R$_\odot$. 
For the binary we used the parameters of Molnar et al.~(\cite{molnar2017}) 
and the Gaia distance ($585\pm 10$~pc) to derive the observable fluxes 
from the stellar atmosphere model fluxes. As it is shown in Fig.~\ref{sed}, 
with the main sequence assumption we cannot get any additional support 
for the existence of the third component. Furthermore, we note that the analysis 
of the NIR spectra made by the NASA Infra-Red Telescope Facility by 
Pavlenko et al.~(\cite{pavlenko2018}) also did not show any obvious 
sign of the presence of a third component. 

On the other hand, we see that the Galex NUV flux at the far left is 
significantly above the model flux. This observation led us to test the 
possibility of another third body scenario, by assuming that it is a 
relatively low-mass white dwarf (WD). We see that adding the flux of a 
$T_{\rm eff}=30000$~K, $R=0.016$~R$_\odot$ WD almost cures the 
discrepant position of the NUV flux, but (obviously) leaves the overall 
fine match at all other wavelengths. Nevertheless, the scatter at the 
peak SED regime is still somewhat high, likely because of the larger 
observational errors of some of the fluxes -- the flux variations due 
to eclipses are rather small, and some fluxes result from averaging 
several observations spread through many orbital periods. For 
instance, the outlier status of the Gaia G-band is hard to understand 
at this moment, because it is surely not due to some 
``single measurement / bad phase'' effect. In addition to the 
scatter, an upward trend toward shorter wavelength near the peak flux 
is also observable. This is sign of the overestimation of the reddening. 
Indeed, as discussed in Appendix C, the reddening value of $E(B-V)=0.082$ 
based on the map of Schlafly \& Finkbeiner~(\cite{schlafly2011}) is 
more than a factor of two higher than the one derived by 
Molnar et al.~(\cite{molnar2017}) from the combination of spectroscopic 
and photometric data. With $E(B-V)=0.03$ we can eliminate the trend and 
considerably lower the likelihood of the WD scenario, but (obviously) 
we cannot cure the scatter around the maximum flux.  

%
%
%
\begin{figure}
 \vspace{0pt}
 \centering
 \includegraphics[angle=-90,width=80mm]{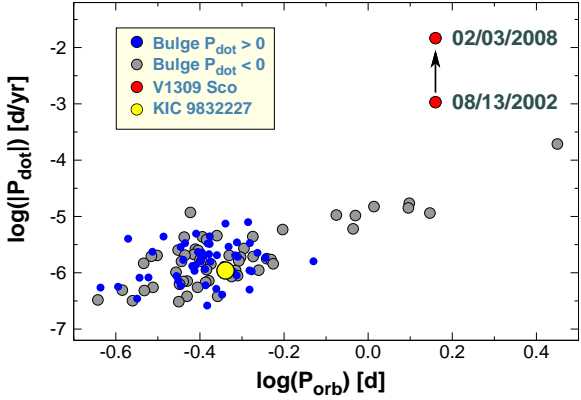}
 \caption{Comparison of the size of the observed period changes of various 
       groups of binaries. The verified pre-nova V1309~Sco stands out with 
       its huge rate and its change within six years of observing time span. 
       On the other hand, KIC~9832227 sits right in the middle of the binaries 
       showing significant, but relatively low, negative period change in the 
       Galactic Bulge sample of Pietrukowicz et al.~(\cite{pietrukowicz2017}). 
       For comparison, we also plot the stars from the above survey, showing 
       period increase.}
\label{pdot_comp}
\end{figure}

Concerning the monotonic component of the total period variation, 
it is interesting to compare the rate of period change with those observed 
in other binaries. As a basis of comparison, we selected the recent 
analysis of $22500$ binaries in the Galactic Bulge by 
Pietrukowicz et al.~(\cite{pietrukowicz2017}). Figure~\ref{pdot_comp} 
shows that the rate of period change of KIC~9832227 is quite similar 
to most of the stars analyzed in the above work. Indeed, V1309~Sco, 
the only verified nova precursor (Tylenda et al.~\cite{tylenda2011})  
clearly stands out with its dramatic period decrease from the other, 
`normal' binaries. Although there is an excess of systems 
with higher negative period change, the nearly equal number of systems 
with positive and negative period changes ($52$ and $56$, respectively) 
indicates that the cause of most of the period changes lies in some 
temporal (long-term) variation rather than an inward spiraling of the 
components. One possibility is the presence of distant perturbers with 
substantially longer period than the time span of the data. In the 
analysis performed by Pietrukowicz et al.~(\cite{pietrukowicz2017}), 
the time base extends from $1997$ to $2015$ -- and, in some cases from 
$1992$, depending on the availability of data from the various OGLE 
phases. Even though this $20$--$25$~yr time span is quite remarkable, 
there are several examples on systems with tertiary components of even 
longer periods (e.g., Sarotsakulchai et al.~\cite{sarotsakulchai2019} 
and Tokovinin~\cite{tokovinin2014} for a review). It is also worthwhile 
to recall that Pietrukowicz et al.~(\cite{pietrukowicz2017}) identified 
also $35$ binaries with cyclic O-C variations. The high number of cyclic 
period changers is further supported by the current survey of the same 
area by Hajdu et al.~(\cite{hajdu2019}), using all available data from 
the OGLE-IV archive. From the $80000$ systems examined, they found 
$992$ with cyclic O-C variations. We conclude that in the specific case 
of KIC~9832227 the total time span of the data used in this investigation 
is $~\sim 19$~years, which leaves still enough room for the 4th-body 
interpretation of the apparent linear period change. 
At the same time, continuation of the monotonic variation of the 
period is also a possibility. This may come from various sources, such 
as mass transfer within the W~UMa components, angular momentum loss 
due to magnetic wind -- e.g., Yang \& Liu~(\cite{yang2003}) and 
Toonen et al.~(\cite{toonen2016}). The complex nature of these processes 
do not allow a more detailed study of these possibilities within the 
limitation of this work. 
 
%
%
\section{Conclusions}
\label{sect5}
The work by Socia et al.~(\cite{socia2018}) on the negation of the 
nova candidate status of KIC~9832227 stimulated our deeper investigation 
of this W~UMa-type binary system. Because the conclusion of 
Socia et al.~(\cite{socia2018}) rests mainly on the misidentification 
of a single early epoch, we thought that the period variation should 
be further investigated by using additional data available from various 
ground- and space-based instruments. Thanks to the wide field survey 
HATNet and to the renewed analysis of the WASP survey, we successfully 
filled out a crucial early period of the O-C variation. In addition 
to the already utilized public domain data by Molnar et al.~(\cite{molnar2017}) 
and Socia et al.~(\cite{socia2018}) we also found extremely useful 
data from the supernova survey ASAS-SN, AAVSO and from the WISE 
satellite. The $39$ epochs were accurate enough to clearly decompose 
the period variation into a linear and a sinusoidal parts. We found 
that the linear rate of change is equal to 
$(-1.097\pm 0.047)\times 10^{-6}$~d/yr. This value fits in the overall rate 
observed in many binary systems. The period and the amplitude of the 
sinusoidal components are equal to $(4925\pm 142)$~d and $(10.98\pm 0.37)$~min.  
We found that the most plausible explanation for the source of this 
variation is the light-time effect of an outer perturber. The 
estimated lower limit of the mass of the outer companion is 
$(0.38\pm 0.02)$~M$\odot$. This value -- assuming that the third 
component is a main sequence star -- is below the upper mass limit 
allowed by the non-detection of the corresponding feature in the 
broadening function analysis of Molnar et al.~(\cite{molnar2017}). 
With the significant deviation of the GALEX NUV point from the 
composite SED constructed under the assumption of a main sequence 
tertiary, we also tested the replacement of this component with 
a white dwarf. We found a better fit, but the contribution of the 
white dwarf component induces a large UV excess at shorter wavelengths. 
Currently, there are no data available at these short wavelengths, 
but future observations might easily confirm or negate the viability 
of the white dwarf scenario. A further ambiguity in the white 
dwarf hypothesis is the poorly known value of the reddening. 
By using $E(B-V)=0.03$ -- i.e., more than a factor of two lower value 
than the one employed in this paper -- decreases the probability of a 
white dwarf companion considerably.

%
\begin{acknowledgements}
We thank Sebastian Otero at AAVSO for answering our questions on
various aspects of the data accessed. We are grateful to the observers
as listed in Appendix A of this paper for their valuable
contribution via the AAVSO database. We acknowledge the improvement 
in the model fit by following the suggestion of the referee of using uniform 
timebase (i.e., BJD) for all data.  
Thanks are due also to Yakiv Pavlenko, for the correspondence on the 
analysis of the NIR spectra. This research has made use of the International 
Variable Star Index (VSX) database, operated at AAVSO, Cambridge, 
Massachusetts, USA. This paper includes data collected by 
the Kepler mission. Funding for the Kepler mission is provided by the
NASA Science Mission directorate.  This publication makes use of data
products from the Wide-field Infrared Survey Explorer, which is a
joint project of the University of California, Los Angeles, and the
Jet Propulsion Laboratory/California Institute of Technology, funded
by the National Aeronautics and Space Administration. The NSVS data 
have been downloaded from the site of SkyDOT, operated by the University 
of California for the National Nuclear Security Administration of the 
US Department of Energy. This research has made use of the VizieR 
catalogue access tool, CDS, Strasbourg, France (DOI: 10.26093/cds/vizier). 
Supports from the National Research, Development and Innovation Office 
(grants K~129249 and NN~129075) are acknowledged. Data presented in 
this paper are based on observations obtained by the HAT station at 
the Submillimeter Array of SAO, and the HAT station at the Fred Lawrence 
Whipple Observatory of SAO. HATNet observations have been funded by 
NASA grants NNG04GN74G and NNX13AJ15G. 
\end{acknowledgements}

%
%
\bibliographystyle{aa} 

%
\appendix
%
%
\section{Data sources and data handling}
Here we describe some relevant pieces of information concerning the 
data used in the O-C analysis as displayed in Table~\ref{o-c_data}. 
First we note that the values for the Vulcan and MLO data have been 
taken from Socia et al.~(\cite{socia2018}) (as a result, all OC values 
are the same for these items). In the case of the AAVSO data, we used 
only {\em V} and {\em I} observations, because other colors did not 
have proper phase coverage. The web sites and the appropriate journal 
references for the data sources are as follows.  
\begin{itemize}
\setlength\itemsep{-1.0em} 
\item
AAVSO: 
\url{https://www.aavso.org/vsx/} \\
\item
ASAS:    
\url{http://www.astrouw.edu.pl/asas/} \\
-- Pojmanski, Pilecki \& Szczygiel~(\cite{pojmanski2005}) and references therein \\
\item
ASAS-SN: \url{https://asas-sn.osu.edu/} \\  
-- Shappee et al.~(\cite{shappee2014}) and Kochanek et al.~(\cite{kochanek2017}) \\
\item
Kepler:  \url{http://keplerebs.villanova.edu/} \\ 
-- Borucki et al.~(\cite{borucki2010}) and Kirk et al.~(\cite{kirk2016}) \\
\item
NSVS:    \url{https://skydot.lanl.gov/nsvs/nsvs.php} \\ 
-- Wo\'zniak et al.~(\cite{wozniak2004}) \\
\item
WASP:    \url{https://wasp.cerit-sc.cz/form} \\ 
-- Pollacco et al.~(\cite{pollacco2006}) \\
\item
WISE:    \url{https://irsa.ipac.caltech.edu/cgi-bin/Gator/nph-dd} \\
\end{itemize} 
The HATNet data are publicly accessible from the website \url{https://data.hatsurveys.org}. 
The data used in this paper are accessible through the CDS link given in Sect.~\ref{sect2}. 
These data result from an early reduction and therefore might be slightly different 
from the data included in the full public HAT archive. Nevertheless, the possible 
differences are expected to have only small effects on the O-C values.

Following the notation of Table~\ref{o-c_data}, the notes below are 
relevant from the point of the analysis presented in this paper (see 
also some related notes in Sect.~\ref{sect2}). 

\noindent
{\em AAVSO-V-1 -- AAVSO-V-5:} Because the data have been gathered from 
different observers with different instruments, they are often not on 
the same zero point. Also, not all data are of the same quality. 
If necessary, we adjusted the quality requirement to the goal of 
segmental sampling by omitting entirely observations owned by a given observer.  
By using the observer assignment code of AAVSO data site, we broke down 
the {\em V} data in the following way. 
{\em AAVSO-V-1:} observer PVEA ({\em Velimir Popov}); 
{\em AAVSO-V-2:} observers SAH (Gerard Samolyk), RIZ ({\em John Ritzel}) 
and TRE ({\em Ray Tomlin}) with a zero point shift applied only to SAH; 
{\em AAVSO-V-3:} observers CDSA ({\em David Conner}), HJW ({\em John Hall}), 
DUBF ({\em Franky Dubois}) and TRE, ZP(CDSA)=+0.037, and all other zero point 
shifts are zero; 
{\em AAVSO-V-4:} observers KTU ({\em Timo Kantola}), HJW and TRE and 
ZP(KTU)$=-0.04$; ZP(HJW)$=-0.025$;  
{\em AAVSO-V-5:} observers TRE, SAH, SNE ({\em Neil Simmons}) and DUBF, 
with ZP(SAH)$=+0.019$, and all other zero point shifts are zero. 

\noindent
{\em AAVSO-I-1:} The data from DUBF and TRE are used, with ZP(DUBF)$=-0.0092$. 

\noindent
{\em ASAS-IV:} Because the data suffer from relative high observational 
noise, the observations made in {\em I} and {\em V} colors have been 
merged with a zero point shift of $+0.717$ for the {\em I} data. We selected 
``grade A'' data from the {\em V} observations and averaged the magnitudes 
in columns \#2 to \#5, corresponding to different aperture sizes. For the 
{\em I} data the ``grade A'' selection resulted only five data points. 
Therefore, we used all data points and averaged only columns \#4 and \#5. 

\noindent
{\em ASAS-SN-1 -- ASAS-SN-4:} The data quality and the number of data points 
allowed us to separate the full dataset into four segments. 

\noindent
{\em HAT-1 -- HAT-5:} To avoid possible signal suppression when the data 
are corrected for instrumental systematics but they are not reconstructed 
by using the full (signal+systematics) model, we used the magnitudes 
derived from an ensemble photometry, based on aperture fluxes. Because of the 
large field of view, the ensemble value of the stars observed during each 
exposure depends on the chip position (e.g., vignetting causes large changes 
in the stellar fluxes). From the three aperture sizes we selected the 
one that resulted the smallest scatter of the folded light curve. Although 
dataset \#4 is already compact in its entirety, it has been divided into 
two parts, because of the large number of data points allows us to do so, 
and thereby effects of short term light curve changes can be assessed.  

\noindent
{\em Kepler-1 -- Kepler-15:} The data quality and the number of data points 
allowed us to separate the full dataset into $15$ segments with $\sim 3500$ 
data points in each segment. We avoided including different Quarters in the 
same data segment, if zero point differences were obvious. Similarly to 
the HAT data, the simple aperture flux (column \#3) is used to avoid 
possible signal degradation when the data are corrected for instrumental 
systematics and not a full signal model is used. 

\noindent
{\em NSVS-12:} We employed the data acquired by the two cameras observing 
the field that hosts KIC~9832227. As in other cases, the moments of 
observations have been transformed to BJD (in this particular case, from MJD).  
A zero point correction of $0.060$~mag is also applied to the dataset 
acquired by one of the cameras. 

\noindent
{\em WASP-1, WASP-3, WASP-4 :} Unfortunately, we had to leave out 
{\em WASP-2}, because of the bad phase coverage of the data points 
for this dataset between HJD 2453901.662 and 2453952.427. 

\noindent
{\em WISE-W12 :} This is a merged set of the data gathered in the W1 and 
W2 filters. The data have been taken in two campaigns, separated by a gap 
of half a year. There is a zero point shift of $\sim 0.03$~mag between 
the two segments. A closer examination also shows that the data from 
the two bands also have zero point differences of the order of 
$\sim 0.01$~mag, that depends also on which of the above 
two campaigns is considered. By performing these campaign- and 
filter-dependent shifts, we ended up with a time series containing $116$ 
items. In spite of the above adjustments, the data still show 
significant scatter of $0.028$~mag around a 4-th order Fourier fit. 
Nevertheless, the accuracy of the O-C value is still acceptable for our 
purpose. 

%
%
\section{The MCMC solution}
In finding the best-fit solution for Eq.~\ref{pol_sin_eq} to the O-C 
data of Table~\ref{o-c_data} (labelled by OC4), we performed a Differential 
Evolution Markov Chain Monte Carlo (DEMCMC) analysis following 
Ter Braak~(\cite{terbraak2006}). We assumed uniform priors on all of 
the parameters given in Eq.~\ref{pol_sin_eq}. Additionally we introduced 
a parameter to account for excess scatter in the O-C values above 
the estimated uncertainties. This scatter was added in quadrature to 
all of the listed errors, and was allowed to vary in the analysis 
using a uniform logarithmic prior. 

\begin{figure}
 \vspace{0pt}
 \centering
 \includegraphics[angle=0,width=80mm]{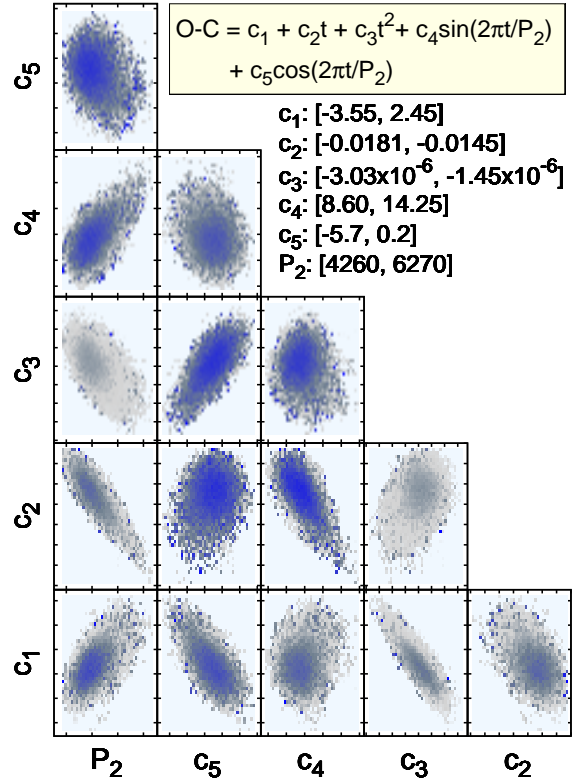}
 \caption{Diagnostic diagram based on the MCMC solution of 
          the fit of the polynomial$+$sinusoidal model (see inset) of 
	  the observed O-C values (see Table~\ref{o-c_data}). Coloring 
	  goes with the $\chi^2$ values: the blue ones have the smallest 
	  $\chi^2$, whereas the gray dots with decreasing intensity are 
	  associated with the less probable solutions. To avoid crowdedness, 
	  we list the parameter ranges of the plots within the figure.}
\label{mcmc}
\end{figure}

The resulting parameter-parameter plots are shown in Fig.~\ref{mcmc}. 
In most cases the cross-correlation effect is not substantial, at 
least in the close neighborhood of the high-density / low-$\chi^2$ 
regimes that determine the most probable solution. We collected the 
final parameters with their $1\sigma$ errors in Table~\ref{mcmc_fit_par}. 
In comparing the solution derived from the standard Least Squares 
fit (see Table~\ref{fit_par}), we see a good agreement for the parameter 
values. However, the MCMC errors are larger. This is because the errors 
include the effect of the various choices of all other parameters, 
whereas in our Least Squares solution the period was fixed to its 
optimum value. 

%
%
%
%

%
%
\begin{table}[!h]
  \caption{Best-fit parameters of the polynomial$+$sinusoidal model by MCMC.}
  \label{mcmc_fit_par}
  \scalebox{1.0}{
  \begin{tabular}{cll}
  \hline\hline
   Coeff & Value & Error  \\ 
 \hline
$c_1$  & $          -0.8765\times 10^{+0} $ & $ 0.5456\times 10^{+0} $ \\ 
$c_2$  & $          -0.1567\times 10^{-1} $ & $ 0.3362\times 10^{-3} $ \\ 
$c_3$  & $          -0.2160\times 10^{-5} $ & $ 0.1424\times 10^{-6} $ \\ 
$c_4$  & $\phantom{-}0.1071\times 10^{+2} $ & $ 0.4788\times 10^{+0} $ \\ 
$c_5$  & $          -0.2539\times 10^{+1} $ & $ 0.5617\times 10^{+0} $ \\ 
\hline
\end{tabular}}
\\
  \scalebox{1.0}{
  \begin{tabular}{llll}
  \multicolumn{4}{c}{Associated quantities} \\
  \hline
  Name         & Unit   &  Value               & Error \\ 
  \hline
  $P_2$        & [d]    &  $\phantom{-}4941  $ & $191$ \\
  RMS          & [min]  &  $\phantom{-}1.382 $ & $0.197$ \\
  {\em Pdot}   & [d/yr] &  $          -1.095\times 10^{-6}$ & $0.072\times 10^{-6}$ \\
  {\em Ampl}   & [min]  &  $          11.007 $ & $0.484$ \\
  \hline
\end{tabular}}
\begin{flushleft}
\begin{minipage}{1.0\linewidth}
{\bf Notes:}{\small \  
For the meaning of the regression coefficients \{$c_i$\} see Fig.~\ref{mcmc}. 
Errors on \{$c_i$\} are {\em full} standard deviations (i.e., including fitting 
error from $P_2$). The RMS of the fit is unbiased, but omitting the correction 
to the 6-th parameter $P_2$. In the MCMC process RMS and {\em Pdot} are 
fitted parameters and the related \{$c_i$\} values are computed from these. 
}
\end{minipage}
\end{flushleft}
\end{table}
%
%

%
%
\section{SED's constituents}
The magnitudes measured for KIC~9832227 have been gathered from the 
VizieR\footnote{\url{https://vizier.u-strasbg.fr/viz-bin/VizieR}} 
site. All magnitudes have been transformed to the Vega system, 
by using the filter zero points as given at the Spanish Visual Observatory 
site\footnote{\url{http://svo2.cab.inta-csic.es/theory/fps/}}. 
To derive dereddened magnitudes, we accepted the reddening given by  
the map of Schlafly \& Finkbeiner~(\cite{schlafly2011}) and accessible 
at the NASA/IPAC Infrared Science 
Archive\footnote{\url{https://irsa.ipac.caltech.edu/applications/DUST/}}. 
This yields: $E(B-V)=0.082\pm 0.002$. Extinction in any given waveband 
has been computed by $A_{\lambda}=R_{\lambda}E(B-V)$, where the extinction 
coefficients $R_{\lambda}$ given by Yuan, Liu \& Xiang~(\cite{yuan2013}), 
Sanders \& Das~(\cite{sanders2018}), Liu \& Janes~(\cite{liu1990}) and  
Davenport et al.~(\cite{davenport2014}) were used. For the XMM-Newton UVW1 band, 
we employed the York Extinction 
Solver\footnote{\url{http://www.cadc-ccda.hia-iha.nrc-cnrc.gc.ca/community/YorkExtinctionSolver/output.cgi}} 
(see McCall~\cite{mccall2004}). All these input data to construct the 
observed SED for KIC~9832227 (see Fig.~\ref{sed}) are listed in 
Table~\ref{sed_tab}. 

We note that the reddening value for KIC~9832227 is far from accurate 
-- in spite of the small formal error bar of the value we used.  
Molnar et al.~(\cite{molnar2017}) compare the spectroscopic and color-calibrated 
$T_{\rm eff}$ values to derive a considerably lower value of $E(B-V)=0.030\pm 0.002$. 
The {\em MWDust} tool by Bovy et al.~(\cite{bovy2016}) yields $E(B-V)=0.021$. 
Exceeding all these values, by using the extinction coefficients of 
Wang \& Chen~(\cite{wang2019}), the Gaia catalog yields $E(B-V)=0.291$ or $0.318$ 
for the $68$\% lower limits, depending if we use their $E(BP-RP)$ or $A_G$ 
values. Although these large values are clearly incompatible with the SED 
based on the accurate distance of the system, it is unclear if some 
considerably lower value than the one we adopted is strongly justifiable 
by the present data.

%
%
\begin{table}[!h]
  \caption{Observed magnitudes of KIC~9832227.}
  \label{sed_tab}
  \scalebox{0.80}{
  \begin{tabular}{lccccccc}
  \hline\hline
  Source  &  Band  &   Band   &  Range    &  Obs  & Vega  &  Error & $R_{\lambda}$ \\
          &        & ($\mu$m) & ($\mu$m)  & (mag) & (mag) &  (mag) & \\
  \hline
GALEX  &  NUV &  0.232 & 0.037 & 17.522 &  15.855 & 0.012 & 7.24 \\     
XMM    & UVW1 &  0.298 & 0.040 & 15.208 &  13.876 & 0.002 & 5.60 \\
M17    &    g &  0.472 & 0.058 & 12.721 &  12.817 & 0.001 & 3.31 \\
M17    &    r &  0.619 & 0.056 & 12.279 &  12.129 & 0.001 & 2.32 \\
M17    &    i &  0.750 & 0.052 & 12.171 &  11.809 & 0.001 & 1.72 \\
M17    &    z &  0.896 & 0.056 & 12.157 &  11.635 & 0.001 & 1.28 \\
GAIA   &    G &  0.674 & 0.210 & 12.293 &  12.293 & 0.005 & 2.39 \\
GAIA   &   BP &  0.532 & 0.117 & 12.626 &  12.626 & 0.015 & 3.17 \\ 
GAIA   &   RP &  0.799 & 0.142 & 11.820 &  11.820 & 0.015 & 1.81 \\
APASS  &    B &  0.431 & 0.042 & 13.081 &  13.023 & 0.121 & 4.10 \\ 
APASS  &    V &  0.545 & 0.044 & 12.351 &  12.305 & 0.087 & 3.10 \\
2MASS  &    J &  1.235 & 0.081 & 11.298 &  11.273 & 0.021 & 0.72 \\
2MASS  &    H &  1.662 & 0.125 & 11.033 &  11.037 & 0.017 & 0.46 \\
2MASS  &    K &  2.159 & 0.131 & 10.998 &  10.983 & 0.016 & 0.31 \\
WISE   &   W1 &  3.353 & 0.331 & 10.921 &  10.921 & 0.023 & 0.19 \\
WISE   &   W2 &  4.603 & 0.521 & 10.946 &  10.946 & 0.021 & 0.15 \\
WISE   &   W3 & 11.561 & 2.753 & 10.865 &  10.865 & 0.072 & 0.30 \\
WISE   &   W4 & 22.088 & 2.051 &  9.492 &   9.492 & 9.999 & 0.08 \\
\hline
\end{tabular}}
\begin{flushleft}
\begin{minipage}{1.0\linewidth}
{\bf Notes:\hskip 1mm}{\small Magnitudes are {\em not} corrected for reddening. 
Further details on items entering in this table can be found in Appendix C.}
\end{minipage}
\end{flushleft}
\end{table}

Transformation to the Vega system has been performed in the following 
way. When the observed magnitudes were given in the AB system (i.e., 
for GALEX, XMM, M17), we computed $m(Vega)=m(AB)+2.5\log(ZP/3631)$, 
where the zero point $ZP$ for the given waveband was taken from the 
Spanish Visual Observatory site. The Gaia and the WISE magnitudes 
are defined on the Vega system\footnote{For Gaia, see the VizieR 
site, for WISE, see: 
\url{http://wise2.ipac.caltech.edu/docs/release/allwise/faq.html}}, 
so their observed and transformed 
values are the same. The APASS magnitudes are in the Johnson 
system, that can be transformed to the AB system based on 
Frei \& Gunn~(\cite{frei1994})\footnote{See also 
\url{http://astroweb.case.edu/ssm/ASTR620/mags.html}}. Then, these 
magnitudes were transformed to the Vega system as already mentioned 
above. The 2MASS magnitudes are nearly on the Vega system, but, 
according to Apellaniz \& Gonzalez~(\cite{apellaniz2018}), there 
might be some systematic differences, yielding the following 
shifts to the published values: $J(Vega)=J(publ)-0.025\pm 0.005$, 
$H(Vega)=H(publ)+0.004\pm 0.005$, $K(Vega)=K(publ)-0.015\pm 0.005$. 
We applied these corrections to the observed 2MASS magnitudes. 

The theoretical SEDs have been downloaded from the Spanish Visual 
Observatory site\footnote{\url{http://svo2.cab.inta-csic.es/theory//newov2/}}, 
using the grid closest to the observed parameters. For the binary 
components we used the $T_{\rm eff}=6000$~K and $5750$~K models, 
both with solar composition, $\log g=4.0$, mixing length and 
microturbulence velocity of $1.25$ and $2$~km/s, respectively. 
These models are without convective overshooting (see Castelli, 
Gratton \& Kurucz~\cite{castelli1997}). The current Gaia distance 
of $585\pm 10$~pc and the radii of Molnar et al.~(\cite{molnar2017}) 
(1.58~R$_\odot$ and 0.83~R$_\odot$ for the primary and secondary 
component, respectively) were used to compute the predicted fluxes. 
In one scenario, the third component was assumed to be a 
$T_{\rm eff}=3500$~K, $R=0.4$~R$_\odot$ main sequence star, whereas 
in another scenario it was assumed to be a white dwarf with 
$T_{\rm eff}=30000$~K, $\log g=8.0$ and $R=0.016$~R$_\odot$. 
In this latter scenario, we used the stellar atmosphere models of 
Koester~(\cite{koester2010}) downloaded from the site cited above. 
The parameters for the white dwarf model are broadly consistent 
with the model values of Romero et al.~(\cite{romero2019}) at 
the temperature chosen, and assuming a mass of $M\sim 0.49$~M$_\odot$. 
This stellar mass is above by $0.11$~M$_\odot$ of our minimum  
value derived from the light time effect (see Sect.~\ref{sect4}).

\end{document}